\documentstyle[12pt,epsfig]{article} 
\def\be{\begin{equation}} 
\def\ee{\end{equation}} 
\newcommand{\bea}{\begin{eqnarray}} 
\newcommand{\eea}{\end{eqnarray}}

\begin{document} 
 
\title{ 
\begin{flushright} 
{\small SMI-16-99} 
\end{flushright} 
\vspace{2cm} 
Noncommutative Gauge Fields on \\ 
Poisson Manifolds 
\footnote{Invited lecture given by I.V. Volovich at the Madeira workshop on 
Noncommutative Infinite Dimensional Analysis, July 1999} 
} 
 
\author{I.Ya. Aref'eva and I.V. Volovich 
\thanks{Permanent address: Steklov Mathematical Institute, 
Gubkin St.8, GSP-1, 117966, Moscow, Russia, 
arefeva@mi.ras.ru volovich@mi.ras.ru} 
\\ $~~~~$ 
\\ 
{\it Centro Vito Volterra} 
\\ 
{\it Universita di Roma Tor Vergata, Rome, Italy}} 
\date {$~$} 
 
\maketitle

\begin {abstract} 
 
It is shown by Connes, Douglas and Schwarz that gauge theory on 
noncommutative torus  describes compactifications of M-theory 
to tori with constant background three-form field. This indicates that 
noncommutative gauge theories on  more general manifolds also can be 
useful in string theory. We discuss a framework to noncommutative 
quantum gauge theory on  Poisson manifolds by using the  deformation 
quantization. The Kontsevich formula for the star product was given 
originally  in terms of the perturbation expansion and it leads to a 
non-renormalizable quantum field theory. We discuss 
the nonperturbative path integral formulation of Cattaneo and Felder 
as a possible approach to construction 
of noncommutative quantum gauge theory on Poisson manifolds. 
Some other aspects of classical and quantum noncommutative field theory 
are also discussed. 
 
\end {abstract} 
 
\newpage 
\section{Introduction} 
 
In a remarkable paper Connes, Douglas and Schwarz \cite{CDS} have shown 
that super-symmetric gauge theory on noncommutative torus is naturally 
related to compactification of Matrix theory \cite{BFSS}. 
This shows that the framework of noncommutative geometry \cite{book,Mad}
can be 
useful in string theory. For some 
reviews and further developments see \cite{DoHu}-\cite{ArArSh}. 
One of the natural questions is 
whether noncommutative gauge theory can be used to describe more general 
compactifications in string theory. In order to have progress in this 
direction  one 
has to develop the theory of noncommutative quantum gauge fields not only 
on the torus but also on  more general manifolds. 
In this note we discuss 
a framework to noncommutative  gauge theory on  Poisson 
manifolds by using the recently developed deformation quantization. 
In particular the question of renormalizability of noncommutative quantum 
theory is discussed. 
 
Gauge theory on {\it noncommutative} torus is equivalent to {\it 
noncommutative} gauge theory on the {\it commutative} (i.e. ordinary) 
torus. Motivated by 
investigations of quantum group symmetries in two dimensional integrable 
models \cite{Tax} 
a proposal of consideration of noncommutative gauge symmetry on 
ordinary manifolds ("quantum group gauge theory") has been suggested in 
\cite{AVqg}, see also \cite{IP,Cas} for further discussions. 
Such a 
theory could be used as an alternative to the conventional Higgs mechanism 
of symmetry breaking. It was also shown \cite{AVn} 
that a noncommutative gauge theory (quantum Boltzmann theory) 
describes the large N limit in QCD. 
Other approaches to field theory on noncommutative 
spaces and harmonic analysis have been discussed in \cite{AJ,Luk}.

A connection of noncommutative geometry with non-Archimedian 
geometry at the 
Planck scale and p-adic mathematical physics \cite{VVZ} has been 
considered in \cite{AV}. 
A proposal to replace the picture of spacetime as a manifold 
to the theory of 
motives with motivic Galois group  as gauge group has been made in 
\cite{mot} as a natural 
extension of p-adic string theory and p-adic gravity \cite{IV,AFDV}. 
In particular L-function of Deligne's motive has been interpreted as the 
partition function of string theory. Applications of motives to quantum 
field theory also have been discussed recently in 
 \cite{Kon2}. A general discussion of relations between number 
 theory and physics is given in \cite{Man}. 
 
Noncommutative gauge theory uses the deformed product (the "star product") 
instead of the ordinary product. The star product \cite{BFD} is a 
generalization of the Moyal product well known in quantum mechanics 
\cite{Ber}. The construction of star product for symplectic manifolds was 
given by De Wilde and Lecomte \cite{Wil} and Fedosov \cite{Fed}. 
Kontsevich \cite{Kon} has found an explicit formula  for the star 
product on Poisson manifolds (nondegenerate Poisson 
structure is reduced to the symplectic structure). 
 His formula is a corollary of a more general result about the 
existence of a quasi-isomorphism between the Hochschild 
 complex of the algebra 
of polynomials and the graded Lie algebra of polyvector fields \cite{Kon}. 
Quantum deformations of the 
Poisson-Lie structures have been considered in \cite{Fad,AAM}.

The formula for the star product was given in \cite{Kon} in terms of a 
perturbation series and if one attempts to use it in noncommutative quantum 
field theory then one gets a non-renormalizable quantum theory. However it 
was pointed out in \cite{Kon} that the formula can be viewed as a 
perturbation series for a topological quantum field theory coupled with 
gravity. Recently Cattaneo and Felder \cite{CF} have shown that 
Kontsevich's 
star product formula is equivalent to the perturbative 
expansion of the path 
integral of a  two-dimensional topological quantum field theory. 
 
In this note we discuss the application of the Cattaneo and Felder 
formulation  to noncommutative gauge theory on Poisson manifolds. 
 
There is also a different aspect of the Cattaneo and Felder 
formulation.  The Moyal bracket gives us a natural way 
to introduce a form-factor to quantum field theories. 
The nonlocality of 
the Moyal bracket is related with random particle 
(one-dimensional world-sheet) and it 
is not enough to make a theory finite. The Cattaneo and Felder 
formula uses a random surface and its generalization to random surfaces 
with non-trivial dynamics in principle could lead to  a finite 
quantum theory. String theory interpretation of the star product has been 
given by Schomerus \cite{Sch}. It would be interesting to explore a 
connection between the star product in deformation 
quantization and Witten's 
product in string field theory \cite{Wit-sft} generalized on curved 
manifolds. 
 
The paper is organized as follows. In the next section 
we will recall some notions of noncommutative geometry, then discuss the 
Moyal product and its applications in quantum field theory. 
In Section 5 the deformation quantization is described and finally in 
Section 6 field theory on Poisson manifolds is discussed. 
 
\section {Noncommutative Geometry and Gauge Fields} 
 
Noncommutative geometry appears in physics in works of the founders 
of quantum mechanics. Heisenberg and Dirac have proposed that 
the phase space of quantum mechanics must be noncommutative 
and it should be described by quantum algebra. After works of von Neumann 
and more recently by Connes mathematical and physical investigations 
in noncommutative geometry became very intensive. 
 
Noncommutative geometry uses a generalization of the known duality 
between a 
space and its algebra of functions, 
see \cite{book,Mad,CDS}. 
If one knows the associative commutative algebra ${\cal A}(M)$ 
of complex-valued functions on topological space $M$ 
then one can restore the space $M$. 
Therefore all topological notions can be expressed in terms of algebraic 
properties of ${\cal A}(M)$.For example, the space of continuous sections of 
vector bundle over $M$ can be regarded as a projective ${\cal A}(M)$-module. 
(We are speaking about left modules. A projective module is a module that can 
be embedded into a free module as a direct summand). 
So vector bundle over compact space $M$ can be identified with projective 
modules over ${\cal A}(M)$. 
 
Now let us make the following generalization. Let ${\cal A}$ 
be an abstract noncommutative algebra which we will interpret as an "algebra 
of functions" on (nonexisting) noncommutative space. 
One can introduce many geometrical notions in this setting. 
For example, a vector bundle is by definition a projective module over 
${\cal A}(M)$ and one can develop a theory of such bundles generalized the 
standard theory. In particular one can define a connection by using the 
following way.  Let ${\cal G}$  be a Lie algebra of derivations of 
 ${\cal A}$ and $\alpha _1,...\alpha _d$ be generators of  ${\cal G}$. 
 If $V$ is a projective module over  ${\cal A}$ (i.e.  a "vector bundle over 
 ${\cal A}$") one defines a connection in $V$ as the set of linear operators 
 $\nabla_1,...\nabla_d$ on $V$ satisfying 
 \be 
 \label{n.1} 
 \nabla_i(a\phi)=a\nabla_i(\phi)+\alpha _i(a)\phi 
 \ee 
 where $a\in  {\cal A}$, $\phi \in V,~~i=1,...d.$ 
 
 In the case when  ${\cal A}$ is an algebra of smooth functions on 
 $R^d$ or on the torus $T^d$ we get the standard notion of connection in a 
 vector bundle. In this case the abelian algebra  ${\cal G}=R^d$ 
 acts on $R^d$ or $T^d$ and correspondingly on ${\cal A}$ 
 by means of translations. The curvature of connection $\nabla_i$ 
\be 
 \label{n.2} 
 F_{ij}=\nabla_i\nabla_j-\nabla_j\nabla_i 
 -f_{ij}^k\nabla_k 
 \ee 
 belongs to the algebra  of 
 endomorphisms of the ${\cal A}$-module $V$. 
 
 The d-dimensional noncommutative torus is defined by its algebra 
 ${\cal A}_\theta$ with generators $U_1,...U_d$ 
 satisfying the relations 
 \be 
 \label{n.3} 
 U_iU_j=e^{2i\pi \theta _{ij}}U_jU_i 
 \ee 
 where $i,j=1,...d$ and $\theta =(\theta _{ij})$ 
 ia a real antisymmetric matrix. 
 The algebra ${\cal A}_\theta$ is equipped with an antilinear involution $*$ 
 obeying $U_i^*=U_i^{-1}$ (i.e. ${\cal A}_\theta$ is a *-algebra). 
 An element of ${\cal A}_\theta$ is a power series 
 \be 
 \label{n.4} 
 f=\sum f(p_1,...p_d)U_1^{p_1}...U_d^{p_d} 
 \ee 
 where $p=(p_1,...p_d)\in Z^d$ and the sequence of complex coefficients 
 $f(p_1,...p_d)$ decreases faster than any power of 
 $|p|=|p_1|+...+|p_d|$ when $|p| \to \infty$. 
 The function $f(p)$ is called the symbol of element $f$. 
 We denote $U^p$ the product $U_1^{p_1}...U_d^{p_d}$. 
 Then one has 
 $U^pU^k=e^{2i\pi \varphi (p,k)}U^{p+k},$ 
 where $\varphi (p,k)=\sum \varphi_{ij}p_ip_j$ 
 and $\varphi_{ij}$ is a matrix obtained from $\theta$ 
after deleting all its elements below the diagonal. 
To simplify the product rule we replace $U^p$ 
by $e^{i\pi \varphi (p,q)}U^p$ so that we have 
 
\be 
 \label{n.5} 
 U^pU^k=e^{i\pi \theta (p,k)}U^{p+k} 
 \ee 
 
If $f$ and $g$  are two elements of ${\cal A}_\theta$,

\be 
 \label{n.6} 
 f=\sum _pf(p)U^{p},~~~ g=\sum _k g(k)U^{k} 
 \ee 
 then the product 
\be 
 \label{n.7} 
 fg=\sum _{p,k}f(p)g(k)U^{p}U^{k}
 \ee 
 $$ 
 =\sum _{p,k}f(p)g(k)e^{2i\pi \theta (p,k)}U^{p+k}=\sum _q(f\star g)(q)U^{q} 
 $$ 
 where the star-product $(f\star g)(q)$ of symbols $f(p)$ 
 and $g(k)$ is

\be 
 \label{n.8} 
 (f\star g)(q)=\sum _{p}f(p)g(q-p)e^{i\pi \theta (p,q-p)} 
 \ee

The differential calculus on the noncommutative torus is introduced by means 
of the derivations $\partial _j$ defined as 
\be 
 \label{n.9} 
 \partial _j U^p=ip_jU^p,~~j=1,...d 
 \ee 
They satisfy the Leibniz rule $\partial _j(fg)= 
\partial _jf \cdot g+ f\cdot \partial _j g$ 
for any $f,g$ $\in {\cal A}_\theta$. 
 
The integral of 
$f=\sum f(p)U^p$ is defined as 
$\int f=f(0)$, which is in correspondence with the commutative case. 
The integral has the property of being the trace on the algebra 
${\cal A}_\theta$, i.e. $\int fg=\int gf$ 
for any $f,g\in {\cal A}_\theta$. Moreover one has 
\be 
 \label{n.10} 
 \int \partial _j f\cdot g=-\int \partial _j g\cdot f 
 \ee 
 
The gauge field $A_i$ on the noncommutative torus is defined as 
 
\be 
\label{g.1} 
A_i=\sum _{p\in Z^d}A_i(p)U^p,~~i=1,...d 
\ee 
Here $A_i(p)$ is a sequence of $N\times N$ complex matrices indexed by a 
spacetime index. It corresponds to the Fourier representation of the ordinary 
gauge theory on commutative torus. 
The gauge field is antihermitian, $A_i^*=-A_i$, 
or $A_i(p)^*=-A_i(-p)$. $A_i$ is an element of a matrix algebra with 
coefficients in ${\cal A}_\theta$ 
and its curvature is defined as 
\be 
\label{g.2} 
F_{ij}=\partial _i A_j-\partial _j A_i+[A_i,A_j] 
\ee 
where $\partial _i$ is a derivative on ${\cal A}_\theta$ 
defined above. The Yang-Mills action is 
\be 
\label{g.3} 
S=-\frac14 \int \mbox{tr} (F_{ij}F^{ij}) 
\ee 
The action is invariant under gauge transformations 
\be 
\label{g.4} 
A_i\to \Omega A_i \Omega ^{-1}+\Omega \partial _i \Omega ^{-1},~~ 
F_{ij}\to \Omega F_{ij}\Omega ^{-1} 
\ee 
where $\Omega$ is a unitary element of the algebra of matrices over 
${\cal A}_\theta$.

\section{Moyal Product} 
 
Let us consider one-dimensional quantum mechanics in the Hilbert space of 
square integrable functions on the real line $L^2({\bf R})$ with ordinary 
canonical operators of position $\hat{q}$ and momenta $\hat{p}$, 
 
\be 
[\hat{p},\hat{q}]=-i\hbar, 
\ee 
acting as $\hat{q}\psi(x)=x\psi(x)$, $\hat{p}\psi(x)=-i \hbar
d\psi(x)/dx$. If a function of two real variables $f(q,p)$ is given 
in terms of its Fourier transform 
 
\be 
f(q,p)=\int e^{i(rq+sp)}\tilde{f}(r,s)dr ds, 
\ee 
then one can associate with it an operator $\hat{f}$ in $L^2({\bf R})$ by 
the following formula 
 
\be 
\hat{f}=\int e^{i(r\hat{q}+s\hat{p})}\tilde{f}(r,s)dr ds. 
\ee 
 
This procedure is called the Weyl quantization and the function $f(q,p)$ 
is called the {\it symbol} of the operator $\hat{f}$ \cite{BS}. 
One has the correspondence 
 
\be 
\hat{f} \longleftrightarrow f=f(q,p) 
\ee 
 
If $\hat{f}^{*}$ is the Hermitian adjoint to $\hat{f}$ then its symbol is 
$f^{*}=f^{*}(q,p)$ 
 
\be 
\hat{f}^{*} \longleftrightarrow \hat{f}^{*}(q,p)=\bar{f}(q,p) 
\ee 
 
If two operators $\hat{f}_1$ and $\hat{f}_2$ are given with symbols 
$f_1(q,p)$ and $f_2(q,p)$ 
then the symbol of product $\hat{f}_1\hat{f}_2$ is given by the 
Moyal product 
$f_1\star f_2=(f_1\star f_2)(p,q)$ as \cite{BS} 
\be 
\label{bs} 
(f_1\star f_2)(p,q)= \sum _{\alpha, \beta} 
\frac{(-1)^{\beta}}{\alpha ! \beta !}(\frac{i\hbar}{2})^{\alpha+\beta} 
(\partial ^\alpha _q \partial ^\beta _p f_1 (p,q)) 
\cdot (\partial ^\beta _q \partial ^\alpha _p f_2 (p,q)) 
\ee 
$$ 
=e^{i\hbar L}(f_1(q_1,p_1)f_2(q_2,p_2))|_{q_1=q_2=q,~~p_1=p_2=p},$$ 
where 
$$ 
L=\frac{1}{2}(\frac{\partial ^2}{\partial q_1 \partial p_2}- 
\frac{\partial ^2}{\partial q_2 \partial p_1}). 
$$ 
This is also can be written as (we set $\hbar =1$) 
\be 
\label{bs2} 
(f_1\star f_2)(p,q)= \frac{1}{(2\pi )^2} 
\int e^{2i[(q-q_2)p_1+(q_1-q)p_2+(q_2-q_1)p]}f_1(q_1,p_1)f_2(q_2,p_2) 
dq_1dq_2dp_1dp_2. 
\ee 
Introducing the constant Poisson structure on the plain 
$\omega ^{\mu\nu}=-\omega ^{\nu \mu}$, $\omega ^{12}=1$ 
and notations $x=(q,p)$, $x_i=(q_i,p_i)$, $i=1,2$ , 
$x_1\omega x_2=q_1p_2-q_2p_1$ 
one writes 
\be 
\label{m.7} 
(f_1\star f_2)(x)=\frac{1}{(2\pi)^2}\int f_1 (x_1)f_2(x_2) 
e^{2i(x\omega x_1+x_1\omega x_2+x_2\omega x)}dx_1dx_2 
\ee 
One also has 
\be 
\label{m.8} 
2i(x\omega x_1+x_1\omega x_2+x_2\omega x)=2i\int _\triangle pdq 
\ee 
where $\triangle$ is the triangle on the plane with vertices $x=(p,q)$, 
$x_1=(p_1,q_1)$ and $x_2=(p_2,q_2)$ and there is the path integral 
representation 
 
\be 
\label{m9} 
e^{2i\int _{\triangle} pdq}=\int e^{2i\int _\triangle pdq}\prod dx(\tau) 
\ee 
One integrate over trajectories $x=x(\tau)$, $0\leq \tau \leq 1$ 
in the phase plane with the boundary conditions 
\be 
\label{m.10} 
x(0)=x(1)=x,~~ 
x(1/3)=x_1,~~x(2/3)=x_2 
\ee 
 
Therefore we obtain that the Moyal bracket is represented in the form 
\be 
\label{e.2} 
(f\star g )(x)=\int K(x,x_1,x_2)f(x_1)g(x_2)dx_1dx_2 
\ee 
where the kernel $K(x,x_1,x_2)$ has a path integral representation 
\be 
\label{e.3} 
K(x,x_1,x_2)=\int e^{iS}\prod dx(\tau) 
\ee 
where 
\be 
\label{e.4} 
S=2\int pdq 
\ee 
and the path integral is taken over the trajectories $x=x(\tau), 
$ $0\leq \tau \leq 1$ subject to the conditions (\ref{m.10}). 
 
\section{Example: Scalar Field} 
To study divergences it is instructive to consider an example of quartic 
interaction of a scalar complex field. We have 
\be 
\label{e.1} 
L=\int d^dx[\frac{1}{2}|\partial _\mu \phi|^2+ 
\frac{m^2}{2}| \phi|^2+\frac{g^2}{4}(\{\bar{\phi}, \phi\}_{M.B.})^2] 
\ee 
 
$$ 
\{\bar{\phi}, \phi\}_{M.B.}=\bar{\phi}\star \phi-\phi\star\bar{\phi} $$ 
$d=2n$. Note that the interaction in (\ref{e.1}) can be also represented 
by using an auxiliary field $\chi$ 
\be 
\label{e.1'} 
L_{int}=g\{\bar{\phi}, \phi\} \chi 
\ee 
We use a relation of the star-product with the Weyl symbol of the product of 
operators 
\be 
\label{e.2a} 
(f\star g )(x)=\int K(x,x_1,x_2)f(x_1)g(x_2)dx_1dx_2 
\ee 
where the kernel $K(x,x_1,x_2)$ has a path integral representation 
\be 
\label{e.3a} 
K(x,x_1,x_2)=\int e^{-\frac{i}{\hbar}S}\prod _{x(0)=x(1)=x,~~ 
x(1/3)=x_1,~~x(2/3)=x_2}dx(\tau) 
\ee 
\be 
\label{e.4a} 
S=\int (x^{1i}dx_{2i}-x^{2i}dx_{1i}) 
\ee 
The integration in (\ref{e.3}) is over loops with three fixed points 
$x(0)=x(1)=x$, $ 
x(1/3)=x_1$, $x(2/3)=x_2$. The choise of points 1/3 and 2/3 on the loop is 
arbitrary and due to reparametrization invariance the result does not depend 
on this choise. 
 
\be 
\label{e.5} 
K(x_1,x_1,x_2) =D(x,x_1)D(x_1,x_2)D(x_2,x) 
\ee 
\be 
\label{e.6} 
D(x,y)=(\frac{1}{2\pi})^{2d/3}\exp \{ \frac{2i}{\hbar}x\omega y\} 
\ee 
and 
\be 
x\omega y= x^{1i}y_{2i}-y^{1i}x_{2i} ,~~ i=1,...n. 
\ee 
 
We interpret (\ref{e.4}) as the propagator of the auxiliary field 
and represent 
(\ref{e.2}) on diagrams as shown in Fig.1. 
\begin{figure}[h] 
\begin{center} 
\epsfig{file=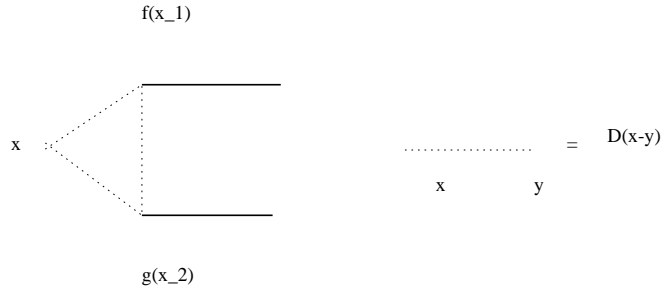, 
   width=250pt, 
   angle=0 
 } 
\end{center} 
\caption{Diagram representation of (\ref{e.2})} 
\label{ff} 
\end{figure} 
 
In these notations the interaction (\ref{e.1'}) has a representation drawn on 
Fig.2. 
\begin{figure}[h] 
\begin{center} 
\epsfig{file=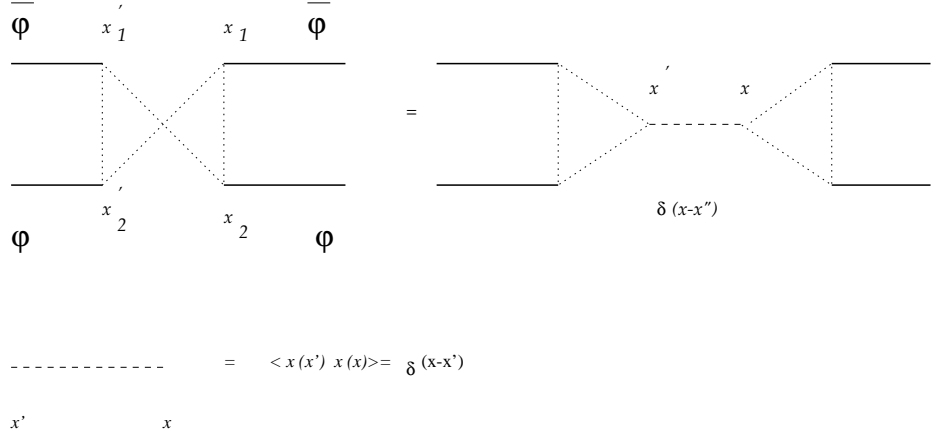, 
   width=350pt, 
   angle=0 
 } 
\end{center} 
\caption{Diagram representation of (\ref{e.1'})} 
\label{4phi} 
\end{figure}

The Fourier representation for $D$-propagator has the form 
\be 
\label{e.7} 
\tilde {D}(p,q)=\int dx dy e^{i(px+qr)}D(x,y)=(\frac{1}{2\pi})^d 
e^{i\frac{\hbar}{2}p\omega q} 
\ee 
and a vertex including auxiliary field $\chi$ can be written in Fourier 
representation as 
\be 
\label{e.8} 
\int \chi (x)(\bar {\phi }\star \phi )(x)dx 
=\int \chi (p)\bar {\phi}(q)\phi (r)\delta (p+q+r)v(p,q,r) 
dpdqdr 
\ee 
$$v(p,q,r)=e^{ip\omega r} $$ 
Fig.2. 
\begin{figure}[h] 
\begin{center} 
\epsfig{file=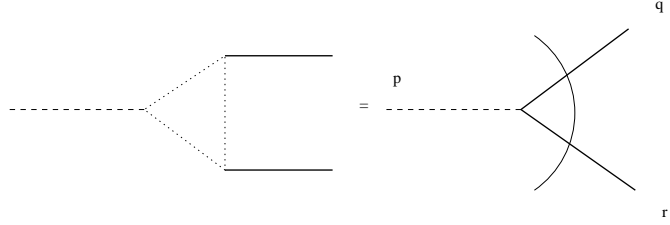, 
   width=250pt, 
   angle=0 
 } 
\end{center} 
\caption{Diagram representation of (\ref{e.8})} 
\end{figure}

\begin{figure}[h] 
\begin{center} 
\epsfig{file=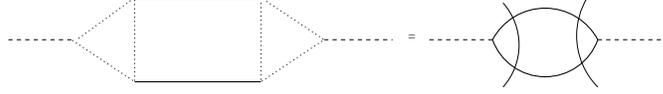, 
   width=250pt, 
   angle=0} 
\end{center} 
\caption{$\chi \chi$-mass renormalization diagram} 
\label{fff} 
\end{figure}

Let us consider the one-loop diagram describing "mass"-renormalization of 
the auxiliary field $\chi$ 
\be 
\label{e.9} 
\int \chi (p)\chi (-p)\Sigma _{\epsilon}(p) 
\ee 
\be 
\label{e.10} 
\Sigma _{\epsilon}(p,\theta)=\Sigma _{\epsilon}(p,0)+ 
\Sigma^+ 
_{\epsilon}(p,\theta)+\Sigma^- _{\epsilon}(p,\theta)= 
\ee 
$$ \frac{1}{2}\int 
\frac{dk}{(k^2+m^2)((p-k)^2)+m^2}[1-e^{2i\theta k\omega p} -e^{-2i\theta 
k\omega p}] 
$$ 
here $d=2n-2\epsilon$. 
Let us prove that the contributions of the last two terms in (\ref{e.10}) 
are finite due to oscillation factors. To show this it is convenient to use 
the standard $\alpha$-representation. 
For example, for the second term in (\ref{e.10}) we have 
Fig.\ref{fff} 
 
\be 
\label{e.11} 
\Sigma^+ _{\epsilon}(p,\theta)=-\frac{(\pi)^{d/2}}{2}\int _0^1dx 
\int _0^\infty \frac{da}{a^{d/2-1}}e^{i[(p^2x(1-x)+m^2)a-\theta ^2p^2/4a]} 
\ee 
For $\theta =0$ we get the standard divergences for $d\geq 4$, 
that can be regularized assuming $d=2n-2\epsilon$, $\epsilon >0$. 
For $\theta \neq 0$, $d=4$ $a=0$ is not a dangerous point. In fact , 
for $d=4$ 
\be 
\label{e.12} 
\int _0^1\frac{da}{a}e^{iB/a}=\int _1^\infty \frac{db}{b^2}e^{iBb} 
<\infty, ~~for ~~B\neq 0 
\ee 
 
Let us note that due to an estimation of the form-factor entering in the 
vertex (\ref{e.8}) 
\be 
|v(p,q,r)|<1 
\ee 
it is evident that the index of 
absolute divergence of a diagram with extra form-factor is the same as 
the index of 
the corresponding diagram with local interaction. 
However this estimation is 
not enough to guarantee the renormalizability of the theory by the 
following  raisons. 
The above estimation do not care of the subdivergences. 
 To have the renormalizability we have to guarantee that all 
 divergencies combine 
to a  special structure to reproduce nonlocal structure of 
interaction. The last problem one meets  already at the 1-loop 
level. As was shown recently in \cite{ren,Sh-ren,Mar} for abelian noncommutative 
Yang-Mills theory 1-loop divergencies combine to renormalizations 
of the wave function and coupling constant and moreover the corresponding 
theory in d=4 is asympotically free. In the next section 
we will describe a diagram technique for perturbation expansion of 
sypersymmetric Yang-Mills theory on noncommutative $T^4$. In the case 
of extended supersymmetries it is enough to study only 1-loop diagrams 
to guarantee the renormalizability.

\section {Noncommutative SYM} 
According to \cite{CDS}, 
the noncommutative $U(1)$ gauge connection can be built by 
\be 
\nabla_i=\partial_i+i\{A_i,\;\;\ \}_{M.B.} 
\ee 
with 
\be 
\{f,g\}_{M.B.}(x)=(f*g)(x)-(g*f)(x)=2i f(x) \sin{(i\pi \theta 
\epsilon^{ij}\stackrel{\leftarrow}{\partial_i}\; 
\stackrel{\rightarrow}{\partial_j})} g(y)|_{y=x}. 
\ee 
The curvature is 
\be 
F_{ij}=[\nabla_{i},\nabla_{j}]=\partial_{[i}A_{j]} + 
i\{A_{i},A_{j}\}_{M.B.} \;\;\; i=0,1,...9 
\ee 
Then $U(1)$ NCYM (on $T^2_{\theta}$) is given by 
\be 
S=\frac{1} {g^2_{YM}}\int d^2x \; F_{ij}F^{ij}. 
\ee 
The above action has the gauge invariance: 
\be 
A_{i}\rightarrow A_{i} +\partial_{i}\epsilon+i\{\epsilon,A_{i}\}_{M.B.} 
\ee 
We can also supersymmetrize this action by adding the correct fermionic and 
scalar degrees of freedom [5,16]. One can take 
 $d=10,\; N=1$ non-Abelian super Yang-Mills theory 
with the group commutators substituted for the 
 Moyal bracket and make 
dimensional reduction to d-dimensional case 
\bea 
S=\frac{1}{ g^2_{YM}}\int d^dx \; F_{\mu\nu}F^{\mu\nu} 
-2 g^2_{YM} (\nabla_{\mu}X^a)(\nabla^{\mu}X^a)+ 2 g^4_{YM}(\{X^a,X^b\}_{M.B.})^2\\ 
-2i \Theta^{\alpha}\Gamma^{\mu}_{\alpha\beta}\nabla_{\mu}\Theta^{\beta} 
+{1\over 4} g_{YM}\Theta^{\alpha}\Gamma^{a}_{\alpha\beta}. 
\{X_a,\Theta^{\beta}\}_{M.B.}, 
\eea 
where $\mu,\nu=0,1,d-1,\; a,b=d,...,9$ and 
$$ 
\nabla_{\mu}X^a =\partial_{\mu}X^a +\{A_{\mu},X^a \}_{M.B.}, 
$$ 
$$ 
\nabla_\mu\Theta^{\alpha}=\partial_\mu\Theta^{\alpha}+ 
\{A_\mu,\Theta^{\alpha}\}_{M.B.}. 
$$ 
A gauge fixed generating functional (in the Lorentz gauge) 
\be 
{\cal Z}[J,\eta,\bar{\eta}]=\int d A dX d\Theta\bar{C}dC 
e^{-S[A,X,\Theta]+S_{GF}[A,X,\Theta]+S_{FP}[A,C,\bar{C}]+sources}. 
\ee 
where 
$C$ and $\bar{C}$ are Faddeev-Popov ghosts, 
$ 
S_{GF}[A_{\mu}]=-\frac{1}{2\alpha}\int (\partial_{\mu}A_{\mu})^{2} 
$ 
and the Faddeev-Popov term is 
$ 
\int \partial_{\mu}\bar{C}(\partial_{\mu}C+g[ A_{\mu},C]) 
$

The  generating functional can be computed 
perturbatively using Feynman diagrams \cite{ren}-\cite{Mar}.
The quadratic terms are  identical to the ones appearing in non abelian 
gauge theories and one gets the following propagators:

$$ 
-\frac{1}{p^{2}}\left( g_{\mu\nu}-(1-\alpha)\frac{p_{\mu}p_{\nu}}{p^{2}}\right) 
\delta(p+q), ~~~~~{\mbox for gauge field} 
$$ 
$$ 
-\frac{1}{p^{2}}\delta(p+q)\delta _{a,b} ~~~~~{\mbox for scalars} 
$$ 
$$ 
-\frac{\slash p}{p^2}\delta(p+q)\delta _{\alpha\beta}~~~~~~{\mbox for fermions} 
$$

Although the propagators are the same as in standard non-abelian 
Yang-Mills theory, 
the interactions take a different form. 
To the three gauge bosons interaction 
we have 
\be 
\label{3g} 
2g\{ 
(p-r)_{\nu}g_{\mu\rho}+(q-p)_{\rho}g_{\mu\nu}+(r-q)_{\mu}g_{\nu\rho}\} 
\sin\theta(p,q)\delta(p+q+r) 
\ee 
 
For the four gauge bosons interaction we have the sum of the following terms 
 
\bea 
\label{4g} 
-4g^{2}\Big( 
&(g_{\mu\rho}g_{\nu\sigma}-g_{\mu\sigma}g_{\nu\rho}) 
\sin\theta(p,q)\sin\theta(r,s)&\nonumber\\ 
+&(g_{\mu\sigma}g_{\nu\rho}-g_{\mu\nu}g_{\rho\sigma}) 
\sin\theta(p,r)\sin\theta(s,q)&\nonumber\\ 
+&(g_{\mu\nu}g_{\rho\sigma}-g_{\mu\rho}g_{\nu\sigma}) 
\sin\theta(p,s)\sin\theta(q,r)& 
\Big)\delta(p+q+r+s). 
\eea 
For the interaction of a gauge boson with ghosts 
we have 
$$ 
2gr_{\mu}\sin\theta(p,q)\delta(p+q+r). 
$$ 
The interaction of scalar fields and gauge fields can been obtained by dimensional 
reduction of the previous vertices. Namely, we can substitute instead of 
$\mu,\nu, \rho, \sigma$ in (\ref{3g}) and (\ref{4g}) 
$m,n,r,s$ and make a reduction 
$$m ~~\to ~~(\mu, a), ~~~n ~~\to ~~(\nu, b)$$ 
$$r ~~\to ~~(\rho, c), ~~~s ~~\to ~~(\sigma, d)$$ 
and now all momenta are d-dimensional. 
The interaction of scalar fields and gauge fields contains two vertices: 
3-vertex describing interaction of gauge fields with scalars 
 
$$2[q_\rho-p_\rho ]g_{bc}\sin\theta(r,p)\delta(p+q+r),$$ 
two scalars - two gauge field vertex 
$$4g^{2} 
g_{\mu\nu}g_{ b a}( 
\sin\theta(p,r)\sin\theta(s,q) 
- 
\sin\theta(p,s)\sin\theta(q,r) 
\Big)\delta(p+q+r+s)$$ 
and 4-vertex describing interaction of  scalars 
\bea 
\label{4s} 
-4g^{2}\Big( 
&(g_{ a\rho}g_{ b d}-g_{ a d}g_{ b\rho}) 
\sin\theta(p,q)\sin\theta(r,s)&\nonumber\\ 
+&(g_{ a d}g_{ b c}-g_{ a b}g_{ c d}) 
\sin\theta(p,r)\sin\theta(s,q)&\nonumber\\ 
+&(g_{ a b}g_{ c d}-g_{ a c}g_{ b d}) 
\sin\theta(p,s)\sin\theta(q,r)& 
\Big)\delta(p+q+r+s). 
\eea

Fermions interact with gauge fields as well as with scalar fields. 
The interaction of a gauge boson with fermions 
is associated with 
$$ 
2gr_{\mu}\sin\theta(p,q)\delta(p+q+r). 
$$ 
 
All these interactions are non local since they involve non polynomial 
functions of the momenta.

It follows from the inequality $|\sin\pi\theta(p,q)|\leq 1$ that 
any diagram which is convergent by powercounting in standard non abelian 
theory is also convergent here. Therefore if we have a local theory 
which has only one-loop divergencies to guarantee renormalizability of its 
noncommutative analogue it is enough to consider only one loop diagrams. 
Explicit calculations similar to \cite{ren}-\cite{Mar} shows that 
noncommutative 
SYM has the same one-loop $\beta$-function as usual SYM 
with $c_2=2$. Therefore for  d=4 extended supersymmetric gauge models 
we get renormalizable theories.

 \section {Deformation Quantization} 
 
Here we describe the Cattaneo and Felder path integral approach \cite{CF} 
to the Kontsevich \cite{Kon} deformation quantization 
\cite {BFD} formula. Let $M$ be a manifold with a Poisson structure on the 
algebra $C^{\infty}(M)$ 
of functions on $M$, 
$\{f,g\}(x)\!=\!\sum_{i,j=1}^d 
p^{ij}(x)\partial_if(x)\partial_jg(x)$ given by 
a skew-symmetric tensor  $p^{ij}$, obeying the 
Jacobi identity 
\be\label{d.1} 
p^{il}\partial_lp^{jk}+ 
p^{jl}\partial_lp^{ki}+ 
p^{kl}\partial_lp^{ij}=0, 
\ee 
The formal deformation quantization is given by means of the star-product, 
i.e. an associative product on $C^\infty (M)[[\hbar ]]$, 
such that for   $f,g \in C^\infty (M)$ 
\be 
\label{d.2} 
f\star g\,(x)=f(x)g(x)+\frac{i\hbar}{2}\,\{f,g\}(x)+O(\hbar^2). 
\ee 
Kontsevich \cite{Kon} gave a formula for the star-product in terms of 
diagrams. The coefficient of $(i\hbar /2)^n$ 
in $f\star g$ is given by a sum of terms labeled by diagram of order $n$. 
To each diagram $\Gamma$ of order $n$ there corresponds a bidifferential 
operator $D_\Gamma$ whose coefficients are differential polynomials, 
homogeneous of degree $n$ in the components $p ^{ij}$ 
of the Poisson structure. 
Kontsevich 's formula is 
\be 
\label{d.3} 
f\star g=fg+\sum_{n=1}^\infty \left(\frac{i\hbar}2\right)^n 
\sum_{\Gamma\, \mathrm{of\, order }\,n}w_\Gamma\, D_\Gamma(f\otimes g). 
\ee 
The weight $w_\Gamma$ is the integral of a differential form 
over the configuration space 
of $n$ ordered points on the upper half plane. 
 
Cattaneo and Felder \cite{CF} have shown that this formula can be interpreted 
as the perturbative expansion of the path integral of a topological sigma 
model. The model has two real bosonic fields $X$ and $a$. $X$ is a map 
from the disc $D=\{u\in R^2,\, |u|\leq 1\}$ to 
$M$ and $a$ is a differential 1-form on $D$ taking values in 
the pull-back by $X$ of the cotangent bundle of $M$. 
In local coordinates, $X$ is given by $d$ functions $X^i(u)$ 
and $a$ by $d$ differential 1-forms $a_i(u)= 
a_{i,\mu}(u)du^\mu$, $i=1,...,d,~\mu=1,2$ 
 
The action reads 
\be 
\label{d.3a} 
S[X,a]=\int_Da_i(u)\wedge dX^i(u)+{\frac12}\,p^{ij} 
(X(u))a_i(u)\wedge a_j(u). 
\ee 
The boundary condition for $a$ is that for 
$u\in \partial D$, $a_i(u)$ vanishes on vectors tangent 
to $\partial D$.

 This model was considered in \cite{Ik}. It is a generalization of a 
 model of two-dimensional gravity with dynamical torsion \cite{KV}, 
 see \cite {Ik}.

The star product is given by the 
semiclassical expansion of the 
path integral 
\be 
\label{d.4} 
f\star g\,(x)=\int 
f(X(1))g(X(0))e^{\frac i\hbar S[X,a]}\prod _{X(\infty)=x} dX\,da. 
\ee 
Here $0$, $1$, $\infty$ are any three cyclically 
ordered points on the unit 
circle. The  path integral is over all 
$X$ and $a$ subject 
to the boundary conditions $X(\infty)=x$, $a(u)(n)=0$ if 
$u\in\partial D$ and $n$ is tangent to $\partial D$. 
 
To evaluate this path integral one has to 
take gauge fixing and renormalization into 
account. This action is invariant under 
the following infinitesimal gauge transformations 
with infinitesimal parameter $\beta_i$, which vanishes on 
the boundary of $D$: 
\be 
\label{d.5} 
\delta_\beta X^i=p^{ij}(X)\beta_j,~~ 
\delta_\beta a_i=-d\beta_i- 
\partial_ip^{jk}(X)a_j\beta_k. 
\ee 
 
The model is quantized \cite {CF} by using the Batalin-Vilkovisky  method.
 To the fields $X^i,a_j$ one adds ghost $\beta _i$, 
antighost $\gamma ^i$ and the scalar Lagrange multiplier $\lambda ^i$ 
together with their antifields $X^+_ 
i,a^{+j}$, $\beta ^{+i}$, $\gamma ^+_i$ and $\lambda ^+_i$ 
with complementary ghost number and degree as differential  forms on $D$. 
The Batalin-Vilkovisky action in a fixed gauge can be written in terms of 
superfields as 
\be 
 \label{d.6} 
S=\int_D d\theta ^2 
\tilde{a}_i D\tilde{X}^i+\frac12\,p^{ij}(\tilde{X})\tilde{a}_i 
\tilde{a}_j-\int _D\lambda ^i\gamma _i^+. 
\ee 
Here $D=\theta^{\mu}\partial/\partial u^{\mu}$, 
\be 
\label{d.7} 
\tilde{X}^i=X^i+\theta^\mu a_\mu^{+i}-\frac12\theta^\mu 
\theta^\nu\beta_{\mu\nu}^{+i}, 
\ee 
\be 
\label{d.8} 
\tilde{a}_i= 
\beta_i+\theta^\mu a_{i,\mu}+\frac12\theta^{\mu} 
\theta^{\nu}X_{i,\mu\nu}^+. 
\ee 
and $\Psi=-\int _D\lambda ^i\gamma _i^+$ is the gauge fixing function 
corresponding to the Lorenz-type gauge $d{*a_i}=0$ (  $*$ is 
the Hodge operator). In the path integral one has to integrate over the 
Lagrangian submanifold defined by equations 
$ 
\phi_\alpha^+=\vec\partial_{\phi^\alpha}\Psi 
$. Then one has $X^+=\beta^+=\lambda^+=0$, 
$\gamma^+_i=d{*a_i}$ and $a^{+i}=*d\gamma^i$. 
The action in component fields is then 
 
\be 
\label{d.10} 
S=\int_Da_i\wedge dX^i+{\frac12}\,p^{ij} 
(X)a_i\wedge a_j-*d\gamma^i\wedge(d\beta_i+ 
\partial_ip^{kl}(X)a_k\beta_l) 
\ee 
$$ 
-\frac14{*d\gamma}^i\wedge {*d\gamma}^j 
\partial_i\partial_jp^{kl}(X)\beta_k\beta_l-\lambda^id{*a_i}. 
$$ 
 
The perturbation expansion was obtained in \cite{CF} 
by taking 
 $X(u)=x+\xi(u)$ with a fluctuation field $\xi(u)$ with 
$\xi(\infty)=0$. The Feynman propagators were deduced from the 
kinetic part 
\be 
\label{d.11} 
S_{0}=\int_Da_i\wedge d\xi^i 
-*d\gamma^i\wedge 
d\beta_i-\lambda^id{*a_i} 
 = 
\int_Da_i\wedge (d\xi^i+*d\lambda^i)+\beta_id{*d}\gamma^i. 
\ee 
of the gauge fixed action. The other 
terms of $S$ are considered as perturbations. 
In terms of superfields 
\be 
\label{d.12} 
S=S_0+S_{int} 
\ee 
where the kinetic part 
\be 
\label{d.13} 
S_0=\int_D\int d^2\theta\,\tilde{a}_j D\tilde{X}^j-\int_D\lambda^i\gamma^+_i, 
\ee 
and 
\be 
\label{d.14} 
S_{int}= 
\frac12\int_D\int d^2\theta\, 
p^{ij}(\tilde{X}(u,\theta)) 
\tilde{a}_{i}(u,\theta)\tilde{a}_{j}(u,\theta). 
\ee 
Here $\tilde{a}_j(u,\theta)=\beta_j(u)+\theta^\mu a_{j,\mu}(u)$, 
$\tilde{\xi}^k(w,\theta)=\xi^k(u)+\theta^\mu a^{+j}_\mu(u)$, with 
$a^{+j}=*d\gamma^j$.

The formula for the star product (\ref{d.3}) is obtained then as the expansion 
of the path integral in powers of $S_{int}$ 
\be 
\label{d.15} 
(f\star g)(x)= \int e^{\frac i\hbar S}{\cal O}=\sum_{n=0}^\infty\frac{i^n}{ 
\hbar^nn!} 
\int e^{\frac i\hbar S_{0}}(S_{int})^n{\cal O}. 
\ee 
where 
\be 
\label{d.16} 
{\cal O}=f(X(1))g(X(0))\delta_x(X(\infty)) 
\ee

\section{Noncommutative Field Theory on Poisson Manifolds} 
 
In this section we discuss noncommutative field theory on Poisson manifolds. 
We would like to use the star product as a generalization of the Moyal 
product to define noncommutative field theory on general Poisson 
manifolds by 
analogy with Sects 4,5. However if one uses the perturbation formula 
(\ref{d.3})
 we immediately run into difficulties because this quantum field 
theory will be nonrenormalizable.  One can see this already on the torus if 
one expands the exponent in Moyal product into the series. Then one gets 
vertices including momenta in higher powers which spoil the 
renormalizability.

Equations of motion for the action (\ref{d.3a}) 
\be 
\label {p.1} 
dX^{i}+p^{ij}(X)da_j =0, 
\ee 
\be 
\label {p.2} 
da_{i}+\frac{1}{2}\partial _ip^{lm}(X)a_l\wedge a_m =0, 
\ee 
have the classical solution $X(u)=x$, $a_i(u)=0$. 
The perturbation expansion from the previous section reproduces the 
Kontsevich formula (\ref{d.3}) however this expansion is not an expansion 
around the classical solution $X(u)=x$. In particular if we apply this 
expansion to the constant Poisson structure then we get a series representing 
the expansion of the Moyal product. 
 
We can obtain the semiclassical expansion if we use the Taylor expansion of 
the Poisson structure 
\be 
\label{p.3} 
p^{ij}(x+\xi (u))=p^{ij}(x)+\partial _k P^{ij}(x)\xi ^k(u)+... 
\ee 
$$ 
=p^{ij}(x)+b^{ij}(x,\xi (u))$$ 
and then set $S=S_0+S_{int}$ where 
\be 
\label{p.4} 
S_0=\int_D\int 
a_i\wedge (d\xi ^i +*d\lambda ^i) 
+\beta _id *d \gamma ^i + \frac{1}{2}p^{ij}(x) a_i\wedge a_j 
\ee 
and 
\be 
\label{p.5} 
S_{int}=\frac12\int \int d^2 \theta b^{ij}(x, \tilde{\xi}(u)) 
\tilde{a}_i\tilde{a}_j 
\ee 
Now expanding the path integral in powers of $S_{int}$ we get the 
semiclassical expansion around the classical solution $X(u)=x$. 
Investigation of this diagram technique and its applications to the field 
theory on Poisson manifolds will be the subject of  a further work.

{\bf Acknowledgments}  This work is
supported in part by INTAS grant 96-0698,  I.Ya.A. 
 is supported also by RFFI-99-01-00166
and I.V.V. by RFFI-99-01-00105 and by grant for the leading
scientific schools. I.V. is grateful to L.Streit for the invitation to
the Madeira workshop on 
Noncommutative Infinite Dimensional Analysis.

\end{document}